\documentclass[12pt]{article}
\usepackage{graphicx}
\usepackage{dcolumn}
\usepackage{bm}

\begin{document}
\begin{titlepage}

\begin{center}
{\Large\bf Charmonium Spectrum and New Observed States} \vskip .5cm

Ailin Zhang\\Department of Physics, Shanghai University, Shanghai,
200444, China\\Email: zhangal@staff.shu.edu.cn
\end{center}

\begin{abstract}
The linearity and parallelism of Regge trajectories is combined with
a hyperfine splitting relation in multiplet to study charmonium
spectrum. It is found that predictions to the spectrum of $1D$
multiplet could be made once another $1D$ state is confirmed. The
newly observed $X(3872)$, $Y(3940)$, $X(3940)$, $Y(4260)$ and
$Z(3930)$ are studied within the charmonium framework.
\end{abstract}
\vskip .5cm



\end{titlepage}

The exploration of hadron spectrum is a central issue in
nonperturbative QCD. Charmonium is the most suitable system to be
studied for its non-relativistic features and a large number of data
accumulated in experiments. It provides people fruitful information
to study the properties of strong interaction. As well known, $q\bar
q$ interaction is described well in terms of potentials including a
color Coulombic $\sim 1/r$ potential, a confinement potential and
some small corrections. However, the exact form of strong
interaction between quark and anti-quark in hadrons is not clear,
the nature of confinement and the relation of the potentials to QCD
are not clear either. All these properties are expected to be
detectable from the hadron spectrum. Furthermore, the study of
charmonium spectrum will be helpful both to identify observed states
and to find new states. Since the discovery of $J/\psi$, many states
have been discovered and identified in charmonium family. In
particular, some new charmonium or charmonium-like states have
lately been observed. This recent achievement in experiments has
stimulated people's great interests on this important field once
again.

$h_c(^1P_1)$ was identified by CLEO\cite{cleo1} in the
isospin-violating reaction
\begin{eqnarray*}
e^+e^-\to\psi(2s)\to\pi^0h_c,h_c\to\gamma\eta_c.
\end{eqnarray*}

$X(3872)$ was first observed  by Belle\cite{belle} in exclusive B
decays
\begin{eqnarray*}
B^\pm\to K^\pm X(3872),X(3872)\to\pi^+\pi^-J/\psi
\end{eqnarray*}
with $M=3872.0\pm 0.6(stat)\pm 0.5(syst)$ MeV and $\Gamma<2.3$
MeV(90\% C.L.). This state is then confirmed by CDF II\cite{cdf},
D0\cite{d0} and BaBar\cite{babar1}.

$Y(3940)$ was observed by Belle\cite{belle1} in exclusive B decays
\begin{eqnarray*}
B\to KY(3940), Y(3940)\to\omega J/\psi.
\end{eqnarray*}
If this enhancement is treated as an S-wave Breit-Wigner resonance,
its mass and total width are $M=3943\pm 11\pm 13$ MeV and
$\Gamma=87\pm 22\pm 26$ MeV.

$X(3940)$ was observed by Belle\cite{belle2} in
\begin{eqnarray*}
e^+e^-\to J/\psi X(3940), X(3940)\to D^\star\bar D
\end{eqnarray*}
with $M=3943\pm 6\pm 6$ MeV and $\Gamma<52$ MeV at $90\%$ C. L.

$Y(4260)$ was observed by BaBar\cite{babar2} in initial-state
radiation events,
\begin{eqnarray*}
e^+e^-\to\gamma_{ISR}Y(4260), Y(4260)\to\pi^+\pi^-J/\psi
\end{eqnarray*}
with $M\sim 4.26$ GeV and $\Gamma\sim 90$ MeV. This state of
$\pi^+\pi^-J/\psi$ was confirmed recently by CLEO
collaboration\cite{cleo2}. The channels $Y(4260)\to\pi^0\pi^0J/\psi$
and $Y(4260)\to K^+K^-J/\psi$ have also been observed in their
study.

$Z(3930)$ was observed in the process $\gamma\gamma\to D\bar D$ by
Belle collaboration\cite{belle3} with $M=3929\pm 5\pm 2$ MeV and
$\Gamma=29\pm 10\pm 2$ MeV, respectively. The states $X(3940)$,
$Y(3940)$ and $Z(3930)$ have just been observed in single experiment
so far and require confirmation by more experiments.

There are many interpretations and suggestions to these new states
since the first announcement of $X(3872)$. For example,
interpretations to the $X(3872)$ could be found in
\cite{x1,x2,x3,x4,x5,x6,x7,x8,x9,x10,x11} and interpretations to the
$Y(4260)$ could be found in \cite{y1,y2,y3,y4,y5,y6,y7}. Some
systemic analyzes to these newly observed states could be found in
\cite{charmonium1,charmonium2,charmonium3}. For the limit of pages,
only parts of literatures are listed here.

The interpretations include conventional charmonium arrangement and
other exotic arrangements outside the $q\bar q$ framework such as
the molecule state, the tetraquark state, the hybrid or the mixing
state among them. However, the identification of these states,
especially the $X(3872)$, is still an open topic.

In order to identify all these states, it is necessary for people to
know well both convenient hadron and exotic hadron. At present time,
people is far away from this. In view of this complexity, we will
study the spectrum of these states within the charmonium framework,
while put aside their exotic interpretations and complicated
production and decay properties.

Hadron spectrum has been studied phenomenologically with Regge
trajectory theory for a long time. Regge trajectory theory indicates
a relation of the square of the hadron masses and the spin of the
hadrons. A Regge trajectory is a line in a Chew-Frautschi\cite{chew}
plot representing the spin of the lightest particles versus their
mass square, t:
\begin{eqnarray}
\alpha(t)=\alpha(0)+\alpha^\prime t
\end{eqnarray}
where intercept $\alpha(0)$ and slope $\alpha^\prime$ depend weekly
on the flavor content of the states lying on corresponding
trajectory. For light quark mesons, $ \alpha^\prime\approx
0.9~GeV^{-2}$.

For radial excited light $q\bar q$ mesons, trajectory on
$(n,M^2)$-plots is described by\cite{ani}
\begin{eqnarray}\label{nm}
M^2=M^2_0+(n-1)\mu^2,
\end{eqnarray}
where $M_0$ is the mass of basic meson, n is the radial quantum
number, and $\mu^2$ (approximately the same for all trajectories)
is the slope parameter of the trajectory.

The behaviors of Regge trajectories in different system, which
indicate that a Regge trajectory is approximately linear while
different trajectories are approximately parallel, have been studied
phenomenologically in many literatures. Regge trajectory with
neighboring mesons (opposite $PC$) stepped by $1$ in $J$ was first
found to be linear and parallel, but was subsequently found to
deviate from linearity and parallelism \cite{nonlinear1,nonlinear2}.
In this case, the exchange degeneracy applies not well. In
phenomenology, the exact deviation depends on peculiar family of
mesons, baryons, glueballs, hybrids and energy regime. In theory,
the non-linearity and the non-parallelism of Regge trajectory result
from intrinsic quark-gluon dynamics which may be flavor and $J$
dependent. Some detailed studies of Regge trajectories could be
found in many more fundamental theories\cite{string}.

In fact, once Regge trajectories with neighboring mesons (same $PC$)
stepped by $2$ in $J$ are under consideration, the linearity and the
parallelism of these trajectories keep well
\cite{ani,nonlinear1,zhangal}, which means that the exchange
degeneracy applies.

Hadron spectroscopy has also been explored in many other
models\cite{model,model1,model2,model3,model4,model5} based on QCD.
In these models, the spectrum of charmonium has been excellently
reproduced for the nonrelativistic features of this system. An
interesting conclusion is that some hyperfine splitting relations
are predicted to exist among the members in a multiplet in potential
models\cite{potential,potential1,potential2}. The S-wave hyperfine
splitting (spin-triplet and spin-singlet splitting), $\Delta
M_{hf}(nS)= M(n^3S_1)-M(n^1S_0)$, is predicted to be finite. For
experimentally observed $M(\psi)$ and $M(\eta_c)$\cite{splitting},
\begin{eqnarray*}
\Delta M_{hf}(1S)&=&M(J/\psi)-M(\eta_c)\simeq 115\pm 2~MeV,
\\
\Delta M_{hf}(2S)&=&M(\psi(2S))-M(\eta_c(2S))\simeq 43\pm 3~MeV.
\end{eqnarray*}

The hyperfine splitting of P-wave or higher L-state is predicted to
be zero
\begin{eqnarray}
\Delta M_{hf}(1P)=<M(1^3P_J)>-M(1^1P_1)\approx 0,\\\nonumber \Delta
M_{hf}(1D)=<M(1^3D_J)>-M(1^1D_2)\approx 0,
\end{eqnarray}
where the deviation from zero is no more than a few MeV. Though the
exact form of potentials may be different in different potential
models, these theoretical predictions are the same. The most
important fact is that the relation in the $1P$ charmonium multiplet
has been proved to be obeyed in a high degree accuracy\cite{pdg}. In
this paper, these relations in the $1P$ and $1D$ multiplets will be
used as facts (or assumptions).

In constituent quark model, $q\bar q$ mesons could be marked by
their quantum numbers, $n^{2S+1}L_J$. For quarkonia, the quantum
numbers $PC$ are determined by $P=(-1)^{L+1}$ and $C=(-1)^{L+S}$.
From PDG\cite{pdg}, we get table~\ref{table-1} for charmonium mesons
without radial excitation. In this table, entries in the first
volume are observed states, entries under $J^{PC}$, $n^{2S+1}L_J$
and mass are confirmed or favored assignment by theoretical analyzes
based on experiments. Entries in the last volume are information
from PDG, and the states marked with a "?" are those not confirmed
and omitted from the summary table. Mass of the most lately
identified $h_c(1P)$ by CLEO\cite{cleo1}($M=3524.4\pm 0.6\pm 0.4$
MeV) is not filled in the table.
\begin{table}
\begin{tabular}{lllllll}
 States & $J^{PC}$ &  $n^{2S+1}L_J$ & Mass(MeV)
& Note\\
\hline\hline $\eta_c(1S)$ & $0^{-+}$ & $1^1S_0$ & 2979.6 & PDG \\
$J/\psi(1S)$ & $1^{--}$ & $1^3S_1$ & 3096.9 & PDG \\
\hline\hline $\chi_{c0}(1P)$& $0^{++}$ & $1^3P_0$ & 3415.2 & PDG \\
$\chi_{c1}(1P)$& $1^{++}$ & $1^3P_1$ & 3510.6 & PDG\\
$h_c$(1P)& $1^{+-}$ & $1^1P_1$ & 3526.2 & PDG ($J^{PC}=?^{??}$)\\
$\chi_{c2}(1P)$& $2^{++}$ & $1^3P_2$ & 3556.3 & PDG\\
\hline\hline $\psi(3770)$& $1^{--}$ & $1^3D_1$ & 3769.9 & PDG \\
$\psi(3836)$ & $2^{--}$ & $1^3D_2$ & $3836\pm 13$ & PDG (?) \\
? & $2^{-+}$ & $1^1D_2$ & ? & ? \\
? & $3^{--}$ & $1^3D_3$ & X(3872)? & $\times$~this work \\
\hline\hline
? & $2^{++}$ & $1^3F_2$ & ? & ? \\
? & $3^{++}$ & $1^3F_3$ & ? & ? \\
? & $3^{+-}$ & $1^1F_3$ & ? & ? \\
? & $4^{++}$ & $1^3F_4$ & ? & ? \\
\hline\hline
\end{tabular}
\caption{Spectrum of charmonium without radial excitation.}
\label{table-1}
\end{table}

Except for the $1^{--}$ $n^3S_1$ states, there is no excited
charmonium having been definitely identified so far. From PDG and
some recent assumptions, we obtain table~\ref{table-2}. In the
table, $Z(3930)$ was suggested as the $\chi_{c2}(2P)$\cite{belle3},
and $Y(3940)$ was suggested as $3^1S_0$\cite{charmonium2} or
$2^3P_0$\cite{gershtein}.
\begin{table}
\begin{tabular}{lllllll}
 States & $J^{PC}$ &  $n^{2S+1}L_J$ & Mass(MeV)
& Note\\
\hline\hline $\eta_c(1S)$ & $0^{-+}$ & $1^1S_0$ & 2979.6
& PDG\\
$\eta_c(2S)$ & $0^{-+}$ & $2^1S_0$ & $3654\pm 6\pm 8$& PDG(?) \\
$\eta_c(3S)$ & $0^{-+}$ & $3^1S_0$ & Y(3940)? & \cite{charmonium2} \\
\hline\hline $J/\psi(1S)$& $1^{--}$ & $1^3S_1$ & 3096.9 & PDG \\
$\psi$(2S)& $1^{--}$ & $2^3S_1$ & 3686.1 & PDG \\
$\psi(4040)$& $1^{--}$ & $3^3S_1$ & $4040\pm 10$ & PDG \\
$\psi(4415)$& $1^{--}$ & $4^3S_1$ & $4415\pm 6$ & PDG \\
\hline\hline $\chi_{c0}(1P)$& $0^{++}$ & $1^3P_0$ & 3415.2 & PDG\\
$\chi_{c0}(2P)$& $0^{++}$ & $2^3P_0$ & Y(3940)? & \cite{gershtein}\\
\hline\hline $\chi_{c1}(1P)$& $1^{++}$ & $1^3P_1$ & 3510.6 & PDG\\
$\chi_{c1}(2P)$& $1^{++}$ & $2^3P_1$ & X(3872)? & ?\\
\hline\hline $h_c(1P)$& $1^{+-}$ & $1^1P_1$ & 3526.2 & PDG ($J^{PC}=?^{??}$)\\
$h_c(2P)$& $1^{+-}$ & $2^1P_1$ & ? & ?\\
\hline\hline$\chi_{c2}(1P)$& $2^{++}$ & $1^3P_2$ & 3556.3 & PDG \\
$\chi_{c2}(2P)$ & $2^{++}$ & $2^3P_2$ & Z(3930) & \cite{belle3} \\
\hline\hline
$\psi(3770)$& $1^{--}$ & $1^3D_1$ & 3770 & PDG \\
$\psi(4160)$ & $1^{--}$ & $2^3D_1$ & $4159\pm 20$ & PDG \\
Y(4260) & $1^{--}$ & $3^3D_1$ & $4260?$ & ? \\
\hline\hline
\end{tabular}
\caption{Spectrum of charmonium with different radial n.}
\label{table-2}
\end{table}

After filling in these two tables, we proceed firstly with the study
of properties of relevant Regge trajectories.

For states without radial excitation, states in each group below
make a trajectory
\begin{eqnarray*}
0^{-+} ~(^1S_0), ~~1^{+-} ~(^1P_1), ~~2^{-+} ~(^1D_2), ~~\cdots ,\\
1^{--} ~(^3S_1), ~~2^{++} ~(^3P_2), ~~3^{--} ~(^3D_3), ~~\cdots ,\\
0^{++} ~(^3P_0), ~~1^{--} ~(^3D_1), ~~2^{++} ~(^3F_2), ~~\cdots ,\\
1^{++} ~(^3P_1), ~~2^{--} ~(^3D_2), ~~3^{++} ~(^3F_3), ~~\cdots .
\end{eqnarray*}
These trajectories were analyzed also in a recent
work\cite{gershtein}.

The high excitation states appeared in these trajectories have not
been observed. In terms of the first two states in each trajectory,
their rough slopes $\alpha^\prime$ are determined
\begin{eqnarray*}
0.282, ~0.327, ~0.392, ~0.419~GeV^{-2},
\end{eqnarray*}
respectively. The slopes of these trajectories increase slowly.
Obviously, these trajectories are not straightly parallel, and the
exchange degeneracy applies not well. These Regge trajectories
indeed fan out as pointed out in reference\cite{nonlinear1}.

For states with radial excitation, the situation is not so
satisfactory for lack of experimental data. At present, there isn't
enough data to make a trajectory even with neighboring mesons
stepped by $1$ in $J$. However, if more states are pinned down, they
will consist of some new Regge trajectories.

Once the $Z(3930)$ is confirmed as the $2^{++}$
$\chi_{c2}(2P)$\cite{belle3}, the $1^{--}$ $2^3S_1$ and the $2^{++}$
$2^3P_2$ will make an excited trajectory. The slope
$\alpha^\prime=0.540~GeV^{-2}$ is larger than the corresponding one
($\alpha^\prime=0.327~GeV^{-2}$) without radial excitation.

If the suggestion of $Y(3940)$ in \cite{gershtein} is right, the
$0^{++}$ $2^3P_0$ and the $1^{--}$ $2^3D_1$ will make an excited
trajectory with slope $\alpha^\prime=0.564~GeV^{-2}$, which is also
larger than the corresponding one ($\alpha^\prime=0.392~GeV^{-2}$)
without radial excitation.

The favorable quantum numbers for $X(3872)$ is now believed to be
$J^{PC}=1^{++}$ or $2^{-+}$\cite{cdf2}. If $X(3872)$ is the
candidate for $2^{-+}$ $2^1D_2$ state, the $0^{-+}$ $2^1S_0$ and the
$2^{-+}$ $2^1D_2$ will make an excited Regge trajectory with slope
$\alpha^\prime=1.219~GeV^{-2}$. If $X(3872)$ is the radial excited
$1^{++}$ $2^3P_1$ charmonium, it makes another trajectory with the
unknown $2^{--}$ $2^3D_2$.

After we have an overall understanding of the properties of Regge
trajectories for charmonium, we start our analyzes to the newly
observed states, and give some predictions to the spectrum of the
$1D$ multiplet.

From previous statements and our analyzes, Regge trajectories with
neighboring mesons stepped by $1$ in $J$ really deviate from
linearity and parallelism. The worst thing is that we don't know how
large the deviations from the linearity and parallelism are for
these trajectories. Though we can give a rough analysis and
predictions to the charmonium spectrum in terms of the trajectory
with neighboring mesons stepped by $1$ in $J$ as did in
\cite{gershtein}, we have no such an intention to go ahead in this
Letter.

In order to make a more precise analysis and potential predictions,
we should make use of the linearity and parallelism for trajectories
with neighboring mesons stepped by $2$ in $J$. Unfortunately, there
exists no Regge trajectory with neighboring mesons stepped by $2$ in
$J$ from experimental data in table~\ref{table-1} and
table~\ref{table-2}. However, if the linearity and parallelism of
Regge trajectories with neighboring mesons stepped by $2$ in $J$
(all mesons with the same $PC$ in one trajectory) is combined with
the hyperfine splitting relation of P-wave or higher L-state
charmonium, some predictions to the spectrum of the $1D$ charmonium
multiplet could be obviously made.

The two lowest Regge trajectories with neighboring mesons stepped by
$2$ in $J$ consist of
\begin{eqnarray*}
0^{-+} ~(^1S_0),~~2^{-+} ~(^1D_2), \\\nonumber 1^{--}
~(^3S_1),~~3^{--} ~(^3D_3).
\end{eqnarray*}
These two trajectories should be parallel and have the same slope.
In these trajectories, it is well known that $0^{-+}$ $\eta_c(1S)$
and $1^{--}$ $\psi(3770)$ have been identified, while $2^{-+}
~1^1D_2$ and $3^{--} ~1^3D_3$ have not been pinned down.

In the $1D$ multiplet, the $1^{--}$ $\psi(3770)$ is identified,
while other three states have not been identified. From our
analyzes, if another $1D$ state ($2^{--}$, $2^{-+}$ or $3^{--}$) is
confirmed, the total spectrum of the $1D$ multiplet could be
predicted.

The favored assignment to $\psi(3836)$ is the $1^3D_2$
state\cite{pdg}, which has not been definitely confirmed. If
$\psi(3836)$ is really confirmed as the $2^{--}$ $1^3D_2$ state, the
masses of $3^{--}$ $1^3D_3$ and $2^{-+}$ $1^1D_2$ could be predicted
as follows.

When the mass of $3^{--}$ $1^3D_3$ is set to be M GeV, the mass of
$2^{-+}$ $1^1D_2$ is found to be $(7M+5\times 3.836+3\times
3.770)/15$ GeV due to zero of hyperfine splitting of the $1D$
charmonium. On the other hand, these two Regge trajectories have the
same slope. Therefore, an equation with M is obtained
\begin{eqnarray}
M^2-3.097^2=({7M+5\times 3.836+3\times 3.770\over 15})^2-2.98^2.
\end{eqnarray}
The solution to this equation is $M=3.981$ GeV. Correspondingly, the
mass of $2^{-+}$ $1^1D_2$ is predicted to be $3.890$ GeV.

Within the charmonium framework, $X(3872)$ was ever interpreted as
the $1^3D_2$ and $1^3D_3$\cite{bettoni}. These arrangements could
also be checked.

If $X(3872)$ is really the $2^{--}$ $1^3D_2$ state, a similar
equation
\begin{eqnarray}
M^2-3.097^2=({7M+5\times 3.872+3\times 3.770\over 15})^2-2.98^2
\end{eqnarray}
with M being the mass of $3^{--}$ $1^3D_3$ is obtained. The masses
of the $3^{--}$ $1^3D_3$ and the $2^{-+}$ $1^1D_2$ are therefore
found to be $4.002$ GeV and $3.912$ GeV, respectively.

If $X(3872)$ is the $3^{--}$ $1^3D_3$, the relations in this $1D$
multiplet do not respect the parallelism of Regge trajectory and
hyperfine splitting relation. The breaking of hyperfine splitting
relation is quite large even though a large deviation from the
parallelism of Regge trajectory is assumed. So we can conclude
safely that this arrangement is impossible and should be ruled out.
This viewpoint is supported by a recent experimental
analysis\cite{cdf2} where the interpretation of $3^{--}$ $1^3D_3$
seems to be excluded.

In any case of these arrangements, once another new state in the
$1D$ multiplet is definitely pinned down, the masses of another two
states will be determined. Obviously, no matter which case is the
reality, it is important to identify another new state in the $1D$
multiplet firstly, and then to find another two states in the
relevant energy regime. Of course, if $X(3872)$ is the $2^{--}$
$1^3D_2$ state, the existed $\psi(3836)$ requires other
interpretation.

In this Letter, the properties of Regge trajectories of charmonium
are studied. We combined the linearity and parallelism of Regge
trajectories with a hyperfine splitting relation, and observed that
some predictions could be given to the spectrum of the $1D$
multiplet. From these analyzes, some results on charmonium spectrum
have been obtained:

1, The assignment of $X(3872)$ as the $3^{--}~1^3D_3$ charmonium
state should be ruled out.

2, If the $X(3872)$ is the $2^{--}$ $1^3D_2$ state, the masses of
the $3^{--}$ $1^3D_3$ and the $2^{-+}$ $1^1D_2$ are predicted to be
$4002$ MeV and $3912$ MeV, respectively.

3, The definite confirmation of $\psi(3836)$ is important for the
$1D$ multiplet. If it is confirmed as the $2^{--}$ $1^3D_2$ state,
the masses of the $3^{--}$ $1^3D_3$ and the $2^{-+}$ $1^1D_2$ is
predicted to be $3981$ MeV and $3890$ MeV, respectively.

These theoretical predictions are expected to give some hints to the
forthcoming experiments. As well known, the linearity and the
parallelism of Regge trajectories with neighboring mesons stepped by
$2$ in $J$ were obtained in the analyzes to many mesons system in
literatures, slight deviations were observed also. The deviations
are expected to affect our conclusions little for their smallness.
However, it will be still interesting to detect how large the
deviations are in charmonium, especially, their relations to the
spin-spin and spin-obit interactions in hadron. The phenomenological
study of the deviations may be important for the study of potential
models.

Acknowledgment: This work is supported in part by National Natural
Science Foundation of China, Foundation of Department of Education
of Shanghai and Shanghai Leading Academic Discipline Project with
project number: T0104.

\end{document}